# Effect of substrate temperature on structural and magnetic properties of c-axis oriented spinel ferrite $Ni_{0.65}Zn_{0.35}Fe_2O_4$ (NZFO) thin films

Dhiren K. Pradhan[1,a)], Shalini Kumari[2,5], Dillip K. Pradhan[3], Ashok Kumar [4], Ram S. Katiyar [5], and R. E. Cohen[1,6,a)]

[1]*Extreme Materials Initiative, Geophysical Laboratory, Carnegie Institution for Science, Washington, DC 20015, USA*
[2]*Department of Physics and Astronomy, West Virginia University, Morgantown, WV 26506, USA*
[3]*Department of Physics and Astronomy, National Institute of Technology, Rourkela-769008, India*
[4]*National Physical Laboratory (CSIR), Delhi-110012, India.*
[5]*Department of Physics and Institute of Functional Nanomaterials, University of Puerto Rico, San Juan, PR 00936,  USA*
[6] *Department of Earth- and Environmental Sciences, Ludwig Maximilians University, Munich 80333, Germany*

## ABSTRACT

Varying the substrate temperature changes structural and magnetic properties of spinel ferrite $Ni_{0.65}Zn_{0.35}Fe_2O_4$ (NZFO) thin films. X-ray diffraction of films grown at different temperature display only (004) reflections, without any secondary peaks, showing growth orientation along the c-axis. We find an increase in crystalline quality of these thin films with the rise of substrate temperature. The surface topography of the thin films grown on various growth temperatures conditions reveal that these films are smooth with low roughness; however, the thin films grown at 800 $^o$C exhibit lowest average and root mean square (rms) roughness among all thin films. We find iron and nickel to be more oxidized (greater $Fe^{3+}$ and $Ni^{3+}$ content) in films grown and annealed at 700 $^o$C and 800 $^o$C, compared to those films grown at lower temperatures. The magnetic moment is observed to increase with an increase of substrate temperature and all thin films possess high saturation magnetization and low coercive field at room temperature. Films grown at 800 $^o$C exhibit a ferrimagnetic–paramagnetic phase transition well above room temperature. The observed large magnetizations with soft magnetic behavior in NZFO thin films above room temperature suggest potential applications in memory, spintronics, and multifunctional devices.

Author to whom correspondence to be addressed. Electronic mail: dhirenkumarp@gmail.com (Dhiren K. Pradhan), rcohen@carnegiescience.edu (R. E. Cohen).

## 1. Introduction

With the rapid development of new devices miniaturizing technology, magnetic structures are being pushed to nanoscale dimensions for potential applications in memory, spintronics, and other multifunctional devices [1-7]. Among the various magnetic materials, the most noteworthy and important materials are strongly correlated spinel oxides of $AB_2O_4$ structure, particularly the spinel ferrites having general formula $AFe_2O_4$ [1-2],[7-10]. Many spinel ferrites exhibit magnetic transition temperature well above room temperature and strong magnetic properties along with high magnetostriction, high resistivity, high dielectric permittivi-ty, good thermal, chemical, and structural stabilities [7-8]. These ferrites also provide a non-zero magnetic moment along with spin-dependent band gaps [7-10]. Ferrimagnetic materials are also widely used as the magnetic candidate in multiferroic composite structures to achieve high magnetization, magnetic phase transition temperatures and strong magnetoelectric (ME) coupling above room temperature [11-15]. Most of the single-phase materials exhibit multiferroicity at cryogenic temperatures and weak magnetoelectric coupling [14-15]. Composites of ferrimagnetic materials with ferroelectric perovskites provide strong ME coupling with both ferroelectric and magnetic transition above room temperature with low leakage current and dielectric loss [14-16]. Nickel-zinc ferrites are the most promising ferrimagnetic materials, possessing high saturation magnetization, low coercive field, high resistivity, high magnetostriction, high dielectric constant, and low dielectric losses with magnetic transition temperature well above room temperature[7-8],[17-19]. Here we study $Ni_{0.65}Zn_{0.35}Fe_2O_4$ (NZFO), since this optimized composition of Ni and Zn shows highest saturation magnetization in the entire Ni-Zn series, considerable magnetostriction ($\lambda_{100} = -24 \times 10^{-6}$, $\lambda_{100} = -13.2 \times 10^{-6}$) and magnetic transition temperatures above 600 K [17-20].

Growing high-quality, single crystalline thin films with atomically smooth surface are a challenge. Many attempts have been made to grow high-quality spinel ferrite thin films by different techniques such as molecular beam epitaxy (MBE), rf sputtering, pulsed laser deposition (PLD), spin-spray, sol-gel, electrochemical deposition, direct liquid phase precipitation and hydrothermal growth [13],[21-26]. The structural and magnetic properties of NZFO are strongly dependent on growth parameters, such as substrate temperature, oxygen partial pressure, substrate to target distance, laser fluence etc [13],[27]. Among all the growth parameters, substrate temperature plays a vital role during growth, but studies on effects of growth temperature on structural and magnetic properties of c-axis oriented NZFO thin films grown by PLD are limited. Here, we have systematically studied the evolution of structural, topography and magnetic properties of NZFO thin films as a function of growth temperature.

## 2. Experimental details

A high purity single phase ceramic target of $Ni_{0.65}Zn_{0.35}Fe_2O_4$ (NZFO) was prepared in air atmosphere by standard solid-state reaction methods; details are reported elsewhere.[12] We fabricated highly oriented NZFO thin films on $LaNiO_3$ (LNO) buffered (LaAlO$_3$)$_{0.3}$ (Sr$_2$AlTaO$_6$)$_{0.7}$ (LSAT) (100) substrates using optimized pulsed laser deposition (PLD) with an excimer laser (KrF, 248 nm). The PLD chamber was initially evacuated to a base pressure of $1.0 \times 10^{-6}$ Torr. Then 60 nm thin LNO buffer layers were grown on LSAT substrates at 700 °C under an oxygen pressure of 200 mTorr. The films were then annealed in a oxygen pressure of 300 Torr for 30 minutes at the same temperature and then slowly cooled to room temperature. The LNO buffer layer used as bottom electrode for electrical measurements. After the deposition of buffer layer, NZFO thin films were deposited at temperatures of 500, 600, 650, 700, 750 and 800 °C under an oxygen pressure of 150 mTorr using a laser density of ~ 1.5 J/cm$^2$ at a deposition frequency of 5 Hz. All the NZFO thin films were then annealed at their respective growth temperature in an oxygen atmosphere of 300 Torr for 30 minutes and later cooled slowly to room temperature. The phase purity and orientation of the NZFO thin films annealed at different temperatures were evaluated by X-ray diffraction (XRD) using a Rigaku Ultima III with Cu $K\alpha_1$ radiation ($\lambda = 1.5405$ Å) operated at a scan rate of 1°/min at room temperature. The film thicknesses were checked using XP-200 profilometer and filmetrics. Room-temperature topography images of NZFO thin films annealed at different temperatures were collected by Atomic force microscopy (AFM) (Veeco) operated in contact mode. High-resolution X-ray photoemission spectroscopy (XPS) was utilized for the elemental analysis of NZFO thin films annealed at different temperatures. Room tempera-ture magnetic hysteresis (M vs H) of all the samples was obtained using a vibrating sample magnetometer (VSM Lakeshore Model 142A). Temperature dependence of magnetic properties of thin films grown and annealed at 800 °C was measured utilizing a PPMS DynaCool (Quantum design) in a wide temperature range of 300-800 K.

## 3. Results

### 3.1 Structural characterization

The phase purity and crystalline quality of all our NZFO thin films grown at different temperature were investigated by high-resolution x-ray diffraction (HRXRD) measurements. Fig. 1(a) shows the XRD patterns of NZFO thin films grown at a different temperature on conducting LaNiO$_3$ (LNO) / (LaAlO$_3$)$_{0.3}$ (Sr$_2$AlTaO$_6$)$_{0.7}$ (LSAT) (100) substrates. These films showed only the (*00c*) (*c* = 4 for NZFO and 1, 2 and 3 for LSAT/LNO) diffraction peaks over a large angle x-ray scans (20° to 80°), confirming that these films are highly c-axis oriented. We did not observe any diffraction peaks from secondary phases in the spectra indicating high purity and high-quality growth of the films. The sharpness of diffraction peaks of NZFO is found to increase with increasing substrate temperature, indicating an increase of crystalline quality of these films [21-23].

### 3.2 Morphological characterization

To investigate the quality of growth, distribution of grains and roughness of all NZFO thin films grown at a different temperatures, surface topography images were captured in contact mode on a scan size of 3 × 3 μm$^2$ (Fig. 2). The surface of all the thin films at different growth temperatures is smooth and homogeneous, free of microcracks, pores or holes. Smooth surfaces with low roughness are necessary for the required physical properties for device applications. We computed the average roughness (*R$_a$*) and root mean square (rms) roughness (*R$_q$*) of each thin film from the atomic force micrographs from the AFM line profiles (Fig. 3). The thin films grown at 500 and 600 °C contain grains of different shape, so the variations in the surface height is more (2.54 to 8.47 nm for film grown at 600 °C), whereas the thin films grown at higher temperatures have densely packed grains of approximately the same shape, hence exhibiting low roughness (*R$_a$* = 0.23 nm for film grown at 800 °C). We did not observe any systematic variation of the roughness with growth temperature. We find that the thin films grown at 800 °C exhibit lowest average and rms roughness among all thin films.

### 3.3 Elemental analysis

We performed high-resolution X-ray photoelectron spectroscopy (XPS) measurements of films grown and annealed at 500, 600, 700, and 800 °C to confirm the presence of all individual elements and measure their oxidation states. Fig. 4 shows the XPS spectra of the film grown at 600 °C. The core-level spectra of Ni 2p, Zn 2p, Fe 2p and O 1s confirm the presence of all individual elements. The electron excitation energies of Ni, Zn, Fe and O observed in these thin films are in good agreement with existing literature, and are characteristics of NZFO [28-29]. The binding energies of all XPS spectra were referenced using the C1s peak (284.6eV). Spin-orbit splitting is observed (2p$_{3/2}$ and 2p$_{1/2}$) are observed for Ni 2p, Zn 2p, and Fe 2p. To probe the effect substrate temperature on the magnetic properties of NZFO thin films, detailed analysis on XPS spectra of Ni and Fe at all four temperatures are shown in Fig. 5. Gaussian-Lorentzian peaks were fitted to the experimental data of Ni 2p and Fe 2p. Ni 2p$_{3/2}$ and Fe 2p$_{3/2}$ peaks were well fitted with double peak Gaussian-Lorentzian profile and we find that Ni and Fe exist within all films as +2 and +3 states for all four temperatures (Fig. 5). Ni $2p_{1/2}$ and Fe $2p_{1/2}$ peaks are well fitted with single peak of Gaussian-Lorentzian profile for 500 °C, 600 °C and a double peak Gaussian-Lorentzian profile for 700°C, 800°C, consistent with +2 valence in the thin films grown at 500 and 600 °C, and a mixture of +2 and +3 states in the thin films grown at 700°C, and 800 °C. The contribution of $Ni^{3+}$ and $Fe^{3+}$ are higher in the NZFO thin films which were annealed at higher temperatures i,e 700°C, and 800°C. A single peak Gaussian-Lorentzian profile for $Zn_{2p}$ was obtained for all four annealing temperatures which confirm that zinc exists within all films as $Zn^{2+}$. Since the amounts of oxidized, and higher moment $Fe^{3+}$ and $Ni^{3+}$ are greater in the thin films grown and annealed at 700°C, and 800°C, higher magnetization is expected in these thin films, and this is observed (Fig. 6(a)). The O 1s can be deconvoluted into two peaks for all four annealing temperatures, where higher binding energy peak belonging to the lattice oxygen and the lower binding energy attributed to the surface adsorbed oxygen. We find no shifting of binding energy, however the full width at half maximum increases with the increase of temperature.

### 3.4 Magnetic properties

We investigated the effects of substrate temperature on magnetic properties by measuring the dc-magnetization as a function of the magnetic field at different temperatures (Fig. 6(a)). We carried out M (H) measurements for the in-plane (magnetic field parallel to the thin film surface) configuration. The diamagnetic response of the substrates, which is significant at higher fields, was subtracted. All of the thin films grown at different growth temperature exhibit perfect saturated magnetic hysteretic behavior above ± 6 kOe, which is a typical signature of ferro/ferrimagnetic behavior. We have estimated saturation magnetization ($M_S$) as a function of annealing temperature (Fig. 6 (a), inset). We find a monotonic increase of $M_S$ with increase of annealing temperature; the observed value at 800°C is around 363 emu/cc. The observed increase in $M_S$ as a function of annealing temperature indicates possible crystallite growth and smoother surface of the films with an increase in temperature. Crystalline growth and decreased surface roughness could also explain the increase in saturation magnetization as a function of substrate/deposition temperature.

We find that the temperature-dependent $H_C$ increases initially with annealing temperature, and reaches a maximum value of 460 Oe at 700 °C, and then drops to ~ 330 Oe with further increase of substrate temperature. This feature can be attributed as expected crossover at 7000C from a single domain to multi-domain behavior with increasing crystallite size. Note that comparatively high $M_S$ and low $H_C$ values of the NZFO thin films indicate the soft magnetic nature.

To probe the magnetic Tc, the temperature dependence of magnetization measurements of NZFO thin films grown at 800 °C have been carried out in the in-plane (magnetic field parallel to the thin film surface) mode. We performed zero-field-cooled (ZFC) and field-cooled (FC) magnetization under a magnetic field of 1000 Oe in a wide temperature range of 300-800 K (Fig. 6 (b)). We observe no bifurcation or hysteresis in M(T ) curve throughout the investigated temperature range. The disappearance of bifurcation in ZFC-FC magnetization curve might be due to the application of field higher than the coercive field [30-31]. We find that magnetization of NZFO thin film decreases slowly up to 690 K and then decreases suddenly with the increase of temperature and vanishes above ~740 K. The estimated ferrimagnetic–paramagnetic transition temperature ($T_C$) is found around 716 (±10) K with a broad transition up to 740 K, which reveals the existence of a short-range spin interaction in the sample above Tc.

## 4. Discussion

The crystalline quality and magnetic moments of NZFO thin films are observed to increase with an increase of substrate temperature. The increase of saturation magnetization with the increase of the substrate temperature is due to better of crystallinity at higher growth temperatures. During the growth, the kinetic energies of ablated particles from the target are not affected by temperature directly, but the density of gas near the substrate decreases as the substrate/growth temperature increases. As a result, the mean free path of the ablated particles might increase. So higher temperatures contribute to the synthesis of high quality crystalline thin films by increasing surface mobility of adatoms [21-23],[27] . The increases of crystalline quality with increase of the substrate temperature of the films are evident from the increase of the XRD peak intensities as shown in Fig.1.. The thin films grown at lower temperature have lower crystalline quality and thus higher surface to volume ratio and as the grain surface is known to be poorly magnetized and results in lower magnetization. In low temperature grown thin films, very small nano-grains are present, which lead to large non-magnetic grain boundary volumes hence low magnetization value. High temperature growth leads to a reduction of the relative grain boundary volume, which causes the increase in the magnetization [21-23]. Liu et al. have reported that NZFO thin films can be fabricated under low temperature, but the magnetic properties of Nickel Zinc ferrite films grown under low temperature were found to be amorphous with high coercivity (Hc) and very low saturation magnetization (Ms) [22-23]. They also concluded that high-temperature post-heating treatment or in-situ heating is needed to obtain a better spinel structure and soft magnetic property with enhanced saturation magnetization. Rajagiri et al. also observed the increase of Ms with an increase of substrate temperature in polycrystalline $MnFe_2O_4$ thin films and the increase of Ms had been attributed to the increase of crystalline quality (crystallite/grain size) with the increase of substrate temperature during the growth [21]. The magnetic materials should exhibit low coercivity, high Ms for spintonics and some other specific application as with very low field magnetization can be switched. Based on our results, NZFO is a potential candidate in this regard. The coexistence of soft magnetic behavior with high Ms and Tc well above room temperature in NZFO thin films makes them suitable for applications in multifunctional magnetic switchable and other spintronic devices [7-8],[10],[13-15],[33].

## 5. Conclusions

Highly c- axis oriented NZFO thin films were grown on LNO buffered LSAT single crystal substrate at different temperature by PLD technique. The crystalline qualities of these thin films were found to increase with an increase of substrate temperature. The surface topography of all the thin films was found very smooth and homogeneous without any microcracks, pores or holes. Observation of photoelectron characteristic peaks of Ni, Zn  Fe, and O in the high-resolution XPS spectra on the surface of all NZFO thin films confirm the presence of all individual elements. The contribution of $Fe^{3+}$ and $Ni^{3+}$ are found to be larger in the thin films grown and annealed at 700 °C and 800 °C compared to the other thin films. These thin films exhibit well saturated hysteretic behavior with high Ms and low Hc at room temperature. The Ms is found to be increased with an increase of substrate temperature, which can be attributed to the increase of crystalline quality with the increase of substrate temperature. The ferrimagnetic- paramagnetic phase transition of the thin films grown at 800°C has been found near to 716  (±10) K well above room temperature. These nanostructures might be useful for nanoscale multifunctional and spintronics device applications as they exhibit well saturated high Ms, soft magnetic behavior, with magnetic Tc well above the room temperature.

## Acknowledgments

This work is partly supported by U.S. Office of Naval Research Grants No. N00014-12-1-1038 and No. N00014- 14-1-0561. D. K. P and R. E. C. are supported by the Carnegie Institution for Science. R. E. C. is also supported by the European Research Council Advanced Grant ToMCaT. R.S. K. acknowledges DoD-AFOSR (Grant #FA9550-16-1-0295).

## References

[1] B.D. Cullity, C.D. Graham, Introduction to magnetic materials, John Wiley & Sons, 2011.

[2] R.A. McCurrie, Ferromagnetic materials: structure and properties, Academic Press, 1994.

[3] H. Zabel, M. Farle, Magnetic nanostructures: spin dynamics and spin transport, Springer, 2012.

[4] S.S. Parkin, M. Hayashi, L. Thomas, Magnetic domain-wall racetrack memory, Science 320 (2008) 190-194.

[5] S. Wolf, D. Awschalom, R. Buhrman, J. Daughton, S. Von Molnar, M. Roukes, A.Y. Chtchelkanova, D. Treger, Spintronics: a spin-based electronics vision for the future, Science 294 (2001) 1488-1495.

[6] R. Sharma, P. Thakur, M. Kumar, P. Sharma, V. Sharma, Nanomaterials for high frequency device and photocatalytic applications: Mg-Zn-Ni ferrites, J. Alloys Compd. 746 (2018) 532-539.

[7] M. Sugimoto, The past, present, and future of ferrites, J. Am. Ceram. Soc. 82 (1999) 269-280.

[8] R. Valenzuela, Novel applications of ferrites, Phys. Res. Int. 2012 (2012).

[9] P. Thakur, R. Sharma, V. Sharma, P. Sharma, Structural and optical properties of $Mn_{0.5}Zn_{0.5}Fe_2O_4$ nano ferrites: Effect of sintering temperature, Mater. Chem. Phys. 193 (2017) 285-289.

[10] S. Emori, D. Yi, S. Crossley, J.J. Wisser, P.P. Balakrishnan, B. Khodadadi, P. Shafer, C. Klewe, A.T. N'Diaye, B.T. Urwin, Ultralow Damping in Nanometer-Thick Epitaxial Spinel Ferrite Thin Films, Nano Lett. (2018), doi: 10.1021/acs.nanolett.8b01261.

[11] K. Jayachandran, J. Guedes, H. Rodrigues, Solutions for maximum coupling in multiferroic magnetoelectric composites by material design, Sci. Rep. 8 (2018) 4866.

[12] J. Ma, J. Hu, Z. Li, C.W. Nan, Recent progress in multiferroic magnetoelectric composites: from bulk to thin films, Adv. Mater. 23 (2011) 1062-1087.

[13] J.A. Moyer, R. Gao, P. Schiffer, L.W. Martin, Epitaxial growth of highly-crystalline spinel ferrite thin films on perovskite substrates for all-oxide devices, Sci. Rep. 5 (2015) 10363.

[14] Y. Wang, J. Hu, Y. Lin, C.-W. Nan, Multiferroic magnetoelectric composite nanostructures, NPG Asia Mater. 2 (2010) 61-68.

[15] C.-W. Nan, M. Bichurin, S. Dong, D. Viehland, G. Srinivasan, Multiferroic magnetoelectric composites: Historical perspective, status, and future directions, J. Appl. Phys. 103 (2008) 1.

[16] D.K. Pradhan, V.S. Puli, S. Kumari, S. Sahoo, P.T. Das, K. Pradhan, D.K. Pradhan, J.F. Scott, R.S. Katiyar, Studies of Phase Transitions and Magnetoelectric Coupling in PFN-CZFO Multiferroic Composites, J. Phys. Chem. C 120 (2016) 1936-1944.

[17] D.K. Pradhan, P. Misra, V.S. Puli, S. Sahoo, D.K. Pradhan, R.S. Katiyar, Studies on structural, dielectric, and transport properties of $Ni_{0.65}Zn_{0.35}Fe_2O_4$, J. Appl. Phys. 115 (2014) 243904.

[18] G.F. Dionne, R.G. West, Magnetic and dielectric properties of the spinel ferrite system $Ni_{0.65}Zn_{0.35}Fe_{2-x}Mn_xO_4$, J. Appl. Phys. 61 (1987) 3868-3870.

[19] T.T. Srinivasan, P. Ravindranathan, L. Cross, R. Roy, R. Newnham, S. Sankar, K. Patil, Studies on high-density nickel zinc ferrite and its magnetic properties using novel hydrazine precursors, J. Appl. Phys. 63 (1988) 3789-3791.

[20] B.P. Rao, A.M. Kumar, K. Rao, Y. Murthy, O. Caltun, I. Dumitru, L. Spinu, Synthesis and magnetic studies of Ni-Zn ferrite nanoparticles, J. Optoelectron. Adv. Mater. 8 (2006) 1703-1705.

[21] P. Rajagiri, B. Sahu, N. Venkataramani, S. Prasad, R. Krishnan, Effect of substrate temperature on magnetic properties of $MnFe_2O_4$ thin films, AIP Advances 8 (2018) 056112.

[22] Y. Liu, Y. Li, H. Zhang, D. Chen, C. Mu, Structural and magnetic properties of NiZn-ferrite thin films prepared by radio frequency magnetron sputtering, J. Appl. Phys. 109 (2011) 07A511.

[23] F. Liu, C. Yang, T. Ren, A. Wang, J. Yu, L. Liu, NiCuZn ferrite thin films grown by a sol-gel method and rapid thermal annealing, J. Magn. Magn. Mater. 309 (2007) 75-79.

[24] S. Kumbhar, M. Mahadik, V. Mohite, K. Rajpure, C. Bhosale, Synthesis and characterization of spray deposited Nickel-Zinc ferrite thin films, Energy Procedia 54 (2014) 599-605.

[25] J. Gunjakar, A. More, K. Gurav, C. Lokhande, Chemical synthesis of spinel nickel ferrite ($NiFe_2O_4$) nano-sheets, Appl. Surf. Sci. 254 (2008) 5844-5848.

[26] D. Pawar, S. Pawar, P. Patil, S. Kolekar, Synthesis of nanocrystalline nickel-zinc ferrite ($Ni_{0.8}Zn_{0.2}Fe_2O_4$) thin films by chemical bath deposition method, J. Alloys Compd. 509 (2011) 3587-3591.

[27] S.-Y. Kim, J.-H. Lee, J.-J. Kim, Y.-W. Heo, Effects of temperature, target/substrate distance, and background pressure on growth of ZnO nanorods by pulsed laser deposition, J. Nanosci. Nanotechnol. 14 (2014) 9020-9024.

[28] J.F. Watts, J. Wolstenholme, An introduction to surface analysis by XPS and AES, Willey, (2003).

[29] C.D. Wagner, Handbook of x-ray photoelectron spectroscopy: a reference book of standard data for use in x-ray photoelectron spectroscopy, Perkin-Elmer, (1979).

[30] P. Joy, P.A. Kumar, S. Date, The relationship between field-cooled and zero-field-cooled susceptibilities of some ordered magnetic systems, J. Phys.: Condens. Matter 10 (1998) 11049.

[31] B. Ghosh, S. Kumar, A. Poddar, C. Mazumdar, S. Banerjee, V. Reddy, A. Gupta, Spin glasslike behavior and magnetic enhancement in nanosized Ni-Zn ferrite system, J. Appl. Phys. 108 (2010) 034307.

[32] S.C. Sahoo, N. Venkataramani, S. Prasad, M. Bohra, R. Krishnan, Substrate Temperature Dependent Anomalous Magnetic Behavior in $CoFe_2O_4$ Thin Film, IEEE Trans. Magn. 47 (2011) 337-340.

[33] R. Sharma, P. Thakur, P. Sharma, V. Sharma, Ferrimagnetic Ni2+ doped Mg-Zn spinel ferrite nanoparticles for high density information storage, J. Alloys Compd. 704 (2017) 7-17.

**Figure Captions**

**Fig. 1.** (Color online) XRD patterns of NZFO thin film grown at different temperatures. The diffraction peaks with symbols * and # corresponds to the peaks of LSAT/LNO (overlapped) and NZFO diffraction patterns, respectively.

**Fig. 2.** (Color online) Atomic force micrographs of NZFO thin films grown at different temperatures. Average roughness *(Rₐ)* and root mean square roughness *(R_q)* of each thin film are listed in their respective growth temperatures.

**Fig. 3.** (Color online) One dimensional AFM line profile which shows the surface height (Z) of NZFO thin films grown at different temperatures within an area of $3 \times 3$ μm². $R_q$ of all thin films are mentioned in their respective growth temperatures.

**Fig. 4.** (Color online) Fitted XPS spectra of Ni ($2p_{1/2}$, $2p_{3/2}$), Zn ($2p_{1/2}$, $2p_{3/2}$), Fe ($2p_{1/2}$, $2p_{3/2}$), O (1s) in NZFO thin film grown at 600 °C which confirms the presence of all elements. The experimental data are represented by black circles and the fitted pattern by the red solid line. The wine colored curve represents the Shirley background.

**Fig. 5.** (Color online) Fitted XPS spectra of Ni ($2p_{1/2}$, $2p_{3/2}$) and Fe ($2p_{1/2}$, $2p_{3/2}$) of NZFO thin films grown at 500, 600, 700, 800 °C. The contribution of $Ni^{3+}$ and $Fe^{3+}$ are observed to be higher in the thin films which were grown at higher temperatures i,e 700°C, and 800°C.

**Fig. 6.** (Color online) (a) Room temperature M-H loops (inset : growth temperature dependence of Ms) of NZFO thin film grown at different temperature (b) Temperature dependence of magnetization measured with zero field cooling (ZFC) and field cooling (FC) with applied static magnetic field of 1000 Oe (300 - 800 K) of NZFO thin film grown at 800 °C. The black solid line represents the ZFC curve and the red solid line represents the FC curve.

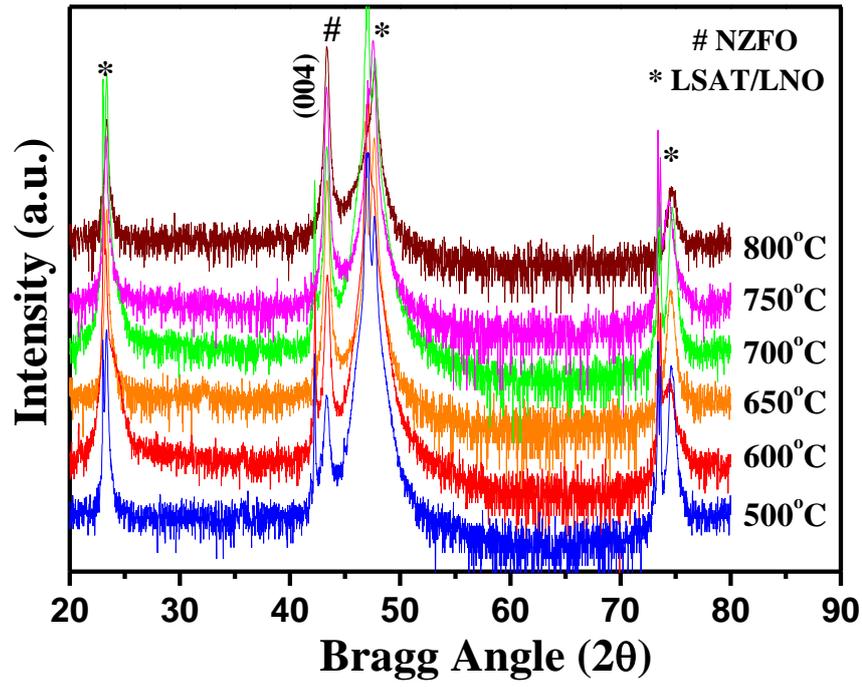

**Figure 1. Pradhan et al.**

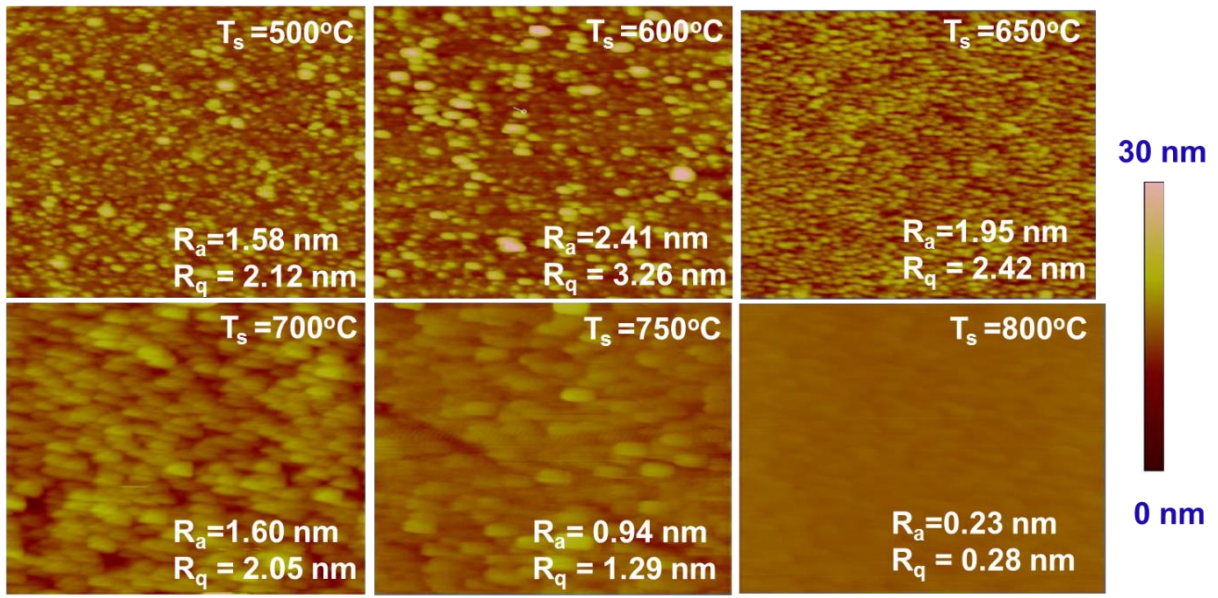

**Figure 2. Pradhan et al.**

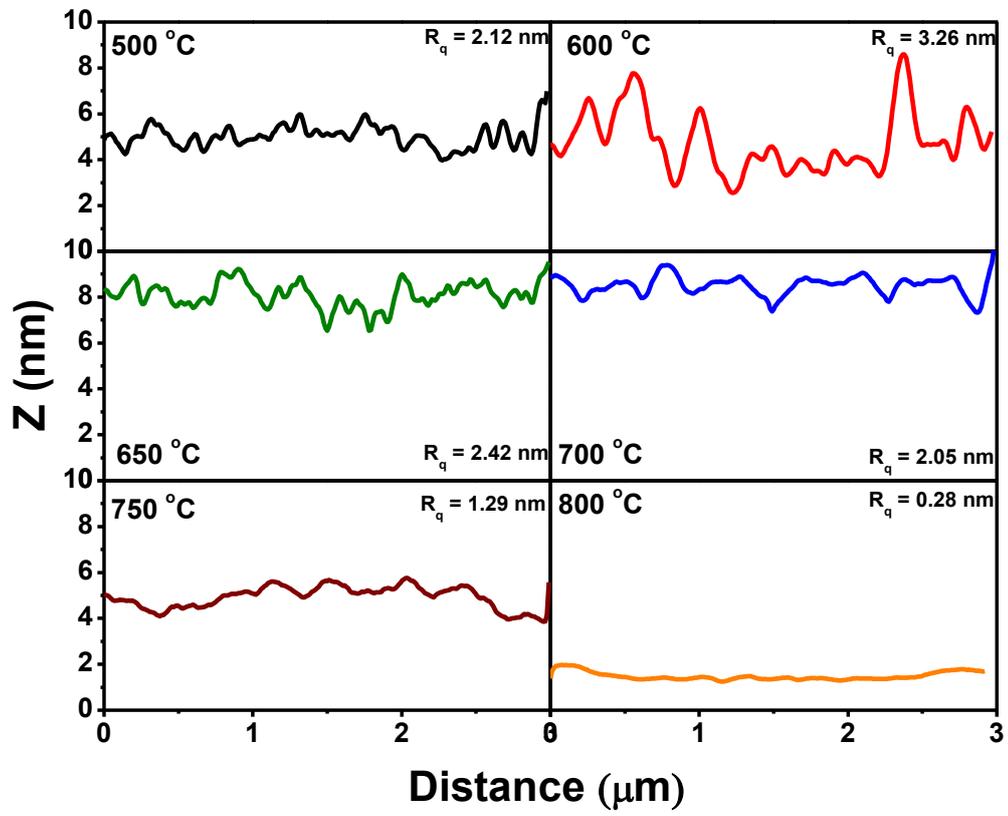

**Figure 3.** Pradhan et al.

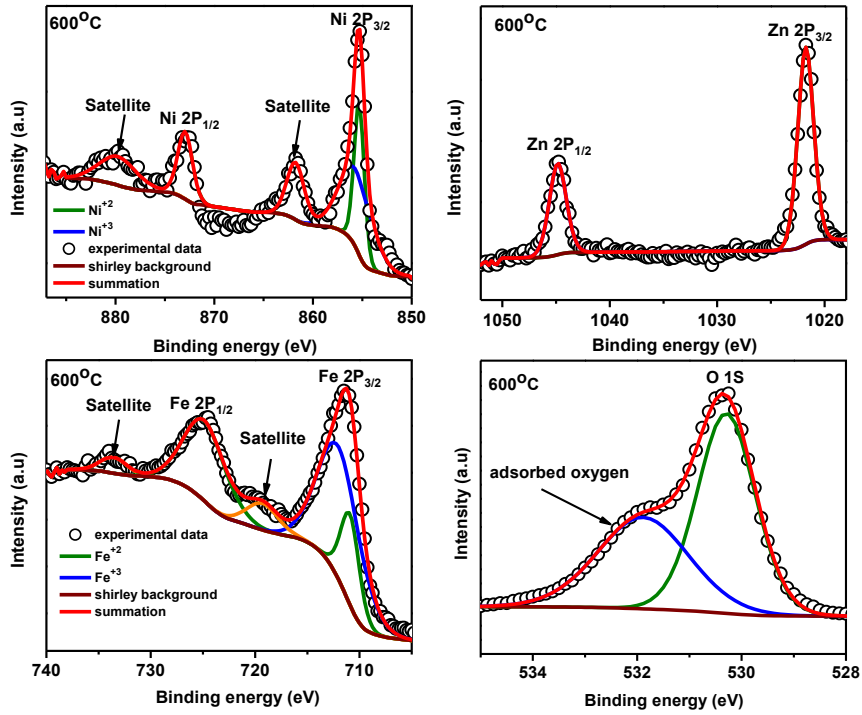

**Figure 4. Pradhan et al.**

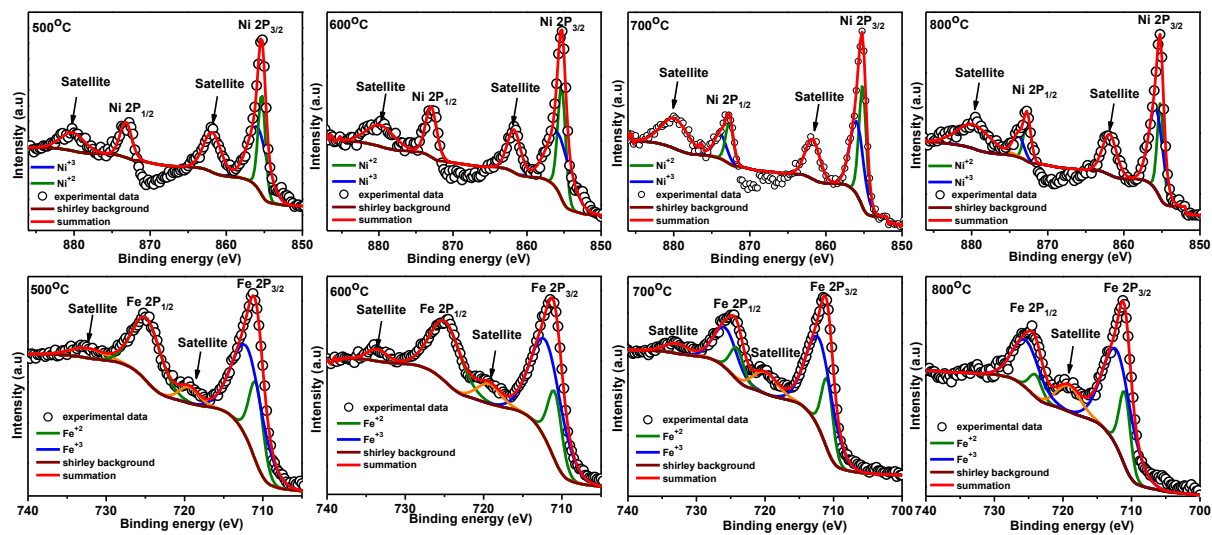

**Figure 5. Pradhan et al.**

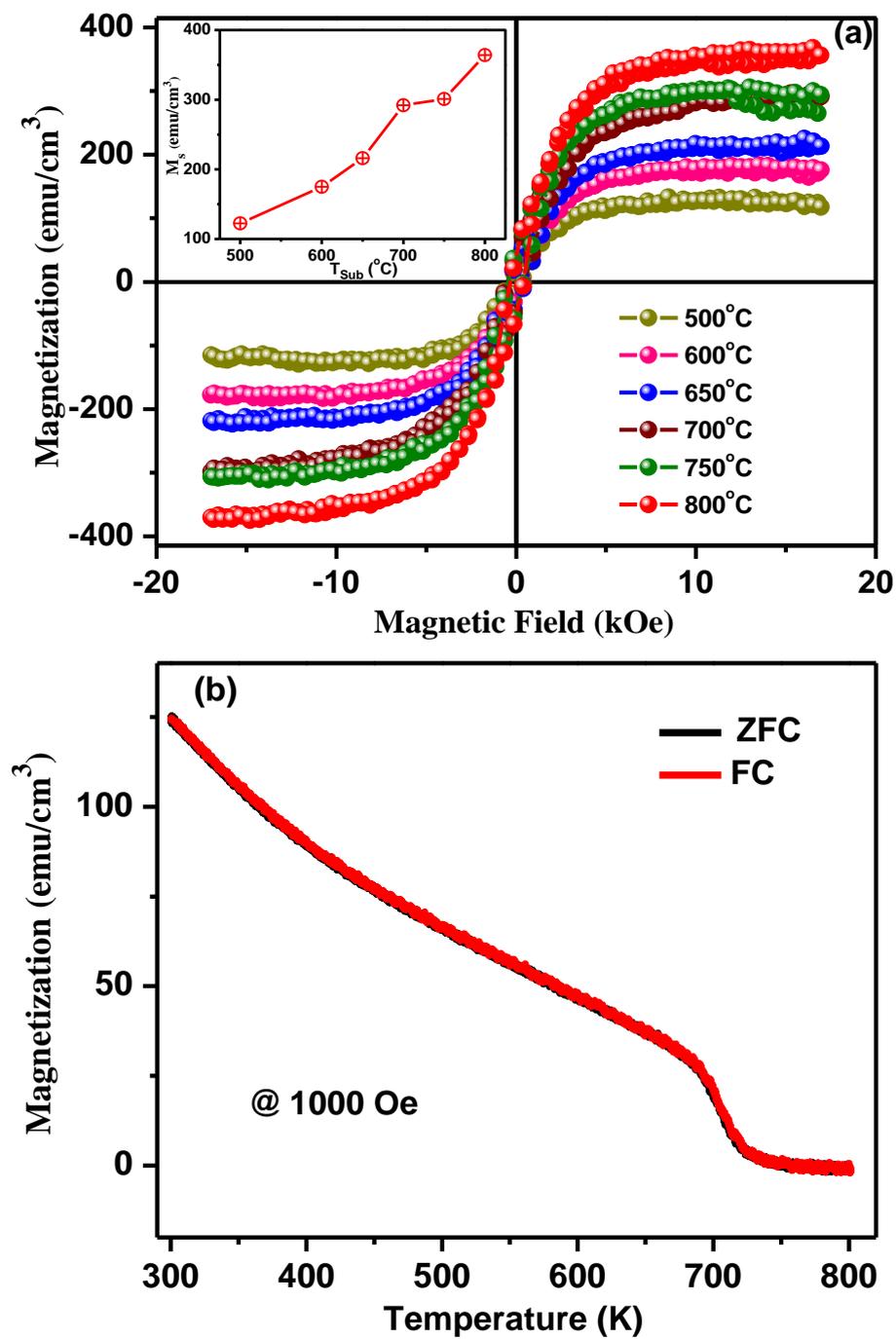

**Figure 6. Pradhan et al.**